# Similar angular THz radiation pattern emitted by filament plasma channel with different length

Andrey V. Koribut, Georgy E. Rizaev, Daria V. Mokrousova, Sergey A. Savinov, Aleksei A. Reutov, Yurii A. Mityagin, Leonid V. Seleznev, and Andrey A. Ionin

**AFFILIATION**

P.N.Lebedev Physical Institute of RAS, Moscow, Russia

**ABSTRACT**

Influence of plasma channel length on angular THz radiation pattern is studied experimentally. It is shown that the angular distribution of THz radiation depends only on the focusing conditions and does not depend on the length of the plasma channel. A qualitative explanation of the screening of THz radiation by a filament plasma is proposed

The plasma channel formed during filamentation of femtosecond laser pulses can serve as a source of terahertz radiation [1-4]. Usually, THz radiation is generated in the so-called two-color filament, where the first and second harmonics of the laser pulse are combined into one filament, and in this case the mechanism for THz emission has been studied and understood. There are much less publications on the generation of THz radiation in a single-color filament, and there is no full understanding of how it occurs. So in [2], the Cherenkov mechanism for THz emission was assumed. In a later work [3], a dipole or quadrupole generation mechanism was proposed. In [4], it was shown that the dipole mechanism makes a decisive contribution to the THz emission. In these works [2-4] it was reported that the angular picture of THz radiation depends on the length of the filament plasma, while the angle of the cone solution is determined by the expression:

$$\theta \sim (\lambda_{THz}/L)^{-1/2}, \qquad (1)$$

the length of the plasma channel L in these works was varied by changing the focusing conditions (the focal length of the focusing element). At the same time, different focusing conditions can lead to a significant (almost an order of magnitude) change in the laser intensity and, as a consequence, to a change in the plasma density by almost three orders of magnitude [5]. Thus, with different focusing, the conditions for the THz emission by the filament plasma will differ significantly. This article is aimed at studying THz radiation from filament plasma at different lengths of the plasma channel under conditions of equal focusing, i.e. at close values of plasma density and intensity in the filament.

In the experiments we use laser pulses generated by a titanium-sapphire laser system. The pulse duration is 90 fs (FWHM), and the central wavelength is $\lambda$ = 740 nm. The diameter of the laser beam is 3.2 mm at a level of 1/e. The pulse energy may be varied by a calibrated diffraction attenuator from 0.1 to 4 mJ. Laser radiation is focused using a lens (Fig. 1). For laser pulses whose power exceeded the critical self-focusing power $P_{cr} = 3.72\lambda^2 / 8\pi n_0 n_2$ (where $n_0$ and $n_2$ are the linear and nonlinear refractive indices, respectively), filamentation and the formation of a plasma channel occur near the focus. As a result of the action of the ponderomotive force and the light pressure, a photocurrent is generated in the plasma, which generates radiation in the terahertz spectral range. This radiation is recorded by a superconducting bolometer operating on the effect of electronic heating in a thin NbN film. The bolometer records THz radiation in the range from 0.3 to 3 THz (in detail in [6]). To obtain the angular distribution of THz radiation, the bolometer may move around a circle having a center in the geometric focus of the lens.

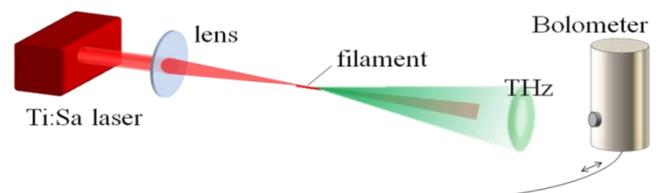

**FIG.1.** Experiment scheme

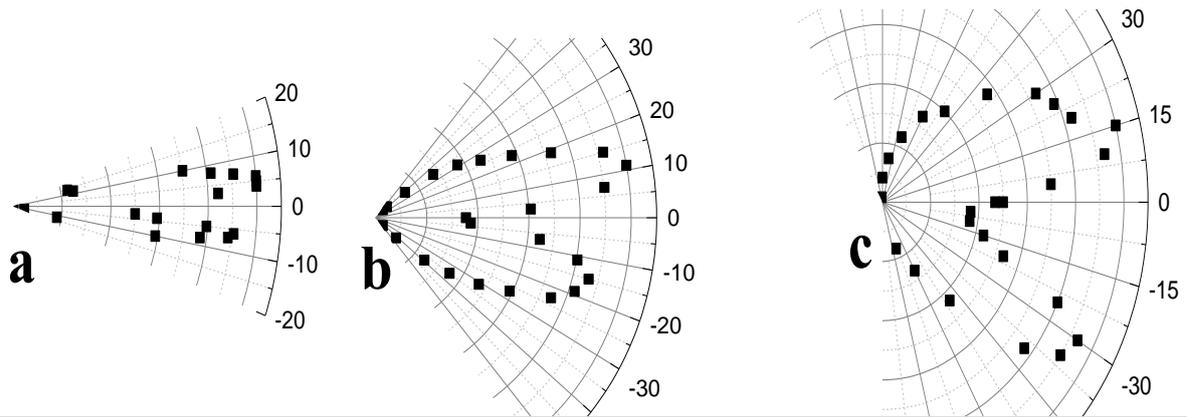

**FIG.2.** The angular distributions of THz emission obtained for lenses with different focal lengths F = 120 cm (a), 25 cm (b) и 5 cm (c).

The angular distributions of THz emission obtained for lenses with different focal lengths are shown in Fig. 2. As in [2], a decrease in the focal length (from 120 cm to 5 cm) leads to a significant increase in the propagation angle of THz emission, which indicates the correctness of the experimental approach.

In the next experiment, the focal length is constant and is 25 cm. Fig. 3. shows the angular distribution of THz radiation for three different laser pulse powers P: <$P_{cr}$ (laser pulse energy 0.25 mJ); ~2.5 $P_{cr}$ (0.8 mJ) and ~13 $P_{cr}$ (4 mJ). There are no significant differences in the angular distributions of THz radiation for laser pulses, whose energy differs by more than an order of magnitude (Fig. 3).

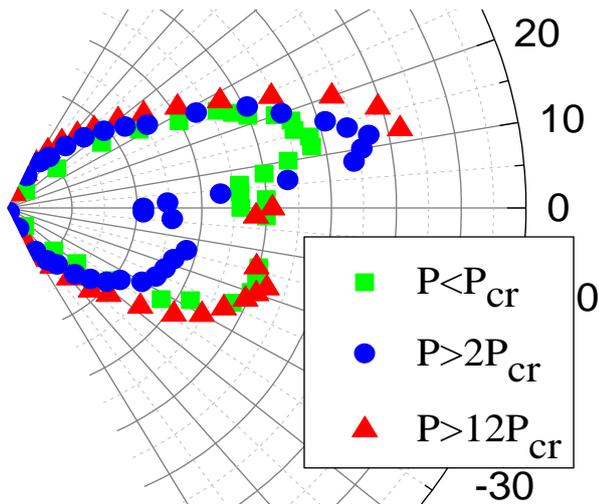

**FIG.3.** Angular distributions of THz radiation for three different laser pulse peak powers.

To estimate the length of plasma channels formed during filamentation, we measure the distribution of linear plasma density along the propagation axis using the capacitor-probe method [7]. In case of subcritical pulse power, we observe a short (~2 cm) plasma channel that is symmetrical with respect to the geometric focus (Fig.4). Increasing the pulse power to 12 $P_{cr}$ elongates the plasma channel to more than 10 cm, while the plasma density increases almost three times, which means an insignificant change in the laser intensity. Based on the formula (1), an increase in the length of the plasma channel by five times should lead to a decrease in the angle $\theta$ by more than two times, that is not observed in the experiment (Fig.3). We conduct similar experiments for the focal length of 120 cm. Laser pulses with energies of 1 and 2.5 mJ form plasma channels with lengths of 30 and 50 cm respectively, while we do not observe any noticeable difference in the direction of THz radiation. The plasma density in these channels differs by about 2 times.

To exclude the influence of differences in plasma density at different energies, we perform an additional experiment with a focal length of 25 cm. The energy of the laser pulse is ~4 mJ. In the experiment, we interrupt the plasma channel, formed during filamentation, by a U-shaped screen (Fig.5a) located at different distances before the linear focus. The form of screen is chosen to eliminate the influence of ablative plasma on the recorded THz signal. The angular distributions of THz radiation obtained without the screen and with the screen located at 1 cm before the focus (corresponds to the vertical line in Fig.4) are shown in Fig.5b. Thus, reducing the length of the plasma channel by almost 2 times (see Fig.4 point -1) at constant plasma densitiy and laser intensity should lead to a 1.4-fold decrease of the angle, according to formula (1), but does not result in noticeable changes in the angular distribution.

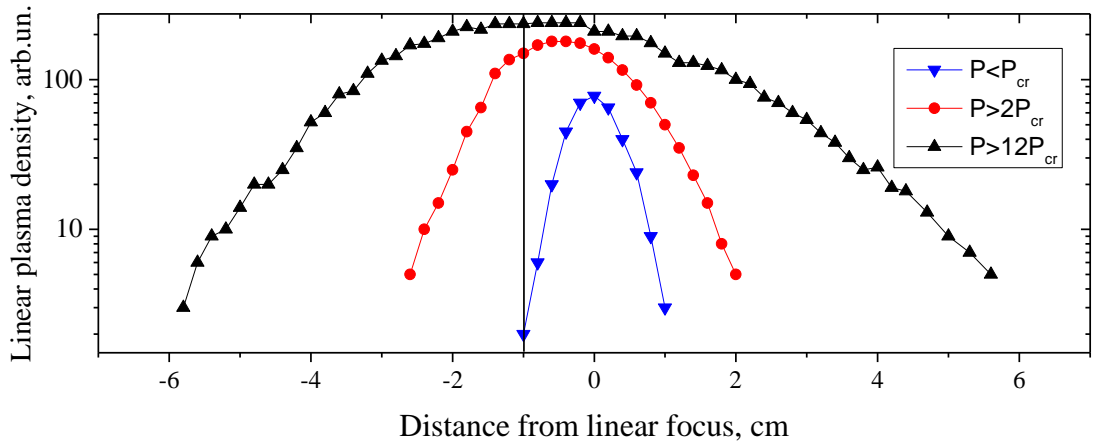

**FIG.4.** Distribution of linear plasma density near the geometric focus at different laser pulse peak powers

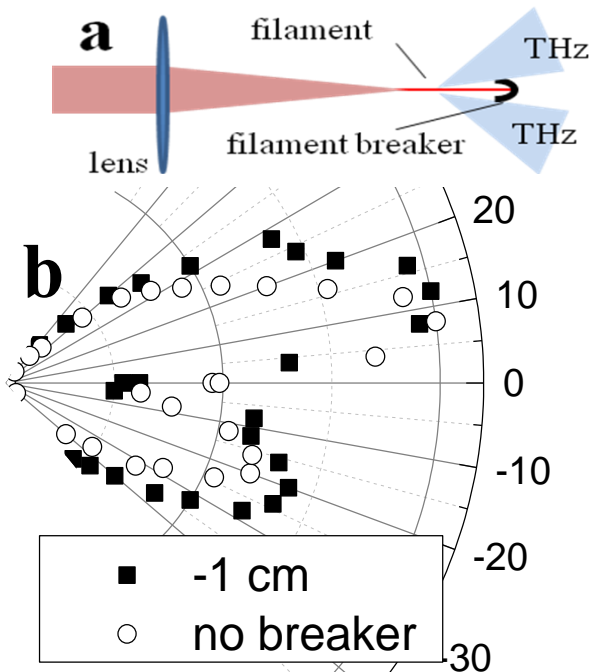

**FIG. 5.** The scheme of interruption of filamentation using a U-shaped screen (a) and the THz angular distribution with and without interruption of filamentation (b).

From the experimental results it is clear that the length of the plasma channel does not play a crucial role in the formation of the angular distribution of THz radiation, in contrast to the focusing conditions. Moreover, as already mentioned, a change in the focusing conditions can change the plasma density in the plasma channel by several orders of magnitude [5]. For electromagnetic radiation with a frequency corresponding to 1 THz, the critical plasma density $\rho_e$, determined by the formula $\rho_e = (m_e\omega^2) / (4\pi e^2)$, is $\sim 10^{16}$ cm$^{-3}$, i.e. even the plasma density characteristic of the collimated filament [1] reaches critical THz radiation. Since THz radiation is generated by the photocurrent of the plasma, the plasma also exists in the immediate vicinity of the source and can screen the THz radiation. Such a screening is schematically shown in Fig. 6. In this case, a significant (by several orders of magnitude) increase in the plasma density should screen THz radiation so that its propagation angle increases significantly (Fig. 6b). In this case, the focusing conditions will play a significant role in the formation of the angular distribution of THz radiation.

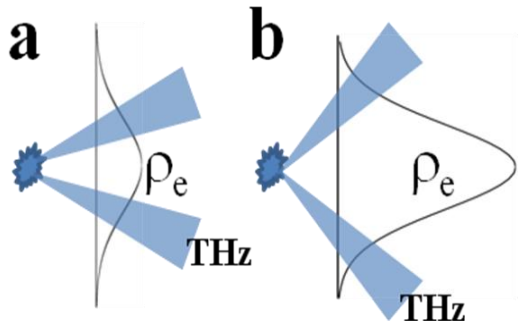

**FIG. 6.** Schematic representation of the screening of THz radiation by filamentation plasma on the axis of propagation with "slight" (a) and "tight" (b) focusing.

As a result of our experiments, it was shown that the length of the plasma channel does not significantly affect the angular distribution of THz radiation generated by the filament plasma photocurrent. It has also been demonstrated that the angular distribution substantially depends on the focusing conditions. A qualitative explanation of the screening of THz radiation by a filament plasma is proposed, which leads to an increase in the propagation angle of THz radiation with an increase in focusing rigidity.

**ACKNOWLEDGMENTS**

This research was supported by the Russian Foundation for Basic Research (RFBR) 20-02-00114.